\documentclass[10pt, 3p,fleqn]{elsarticle}

\makeatletter
\def\ps@pprintTitle{%
 \let\@oddhead\@empty
 \let\@evenhead\@empty
 \def\@oddfoot{\centerline{\thepage}}%
 \let\@evenfoot\@oddfoot}
\makeatother

\usepackage[british]{babel}
\usepackage[utf8]{inputenc}
\usepackage{graphics}
\usepackage{graphicx}
\usepackage{amsmath,amssymb,esint}
\usepackage{lineno}
\usepackage{multirow}
\usepackage{subfig}
\usepackage{natbib}
\usepackage{eucal}
\usepackage{mathtools}
\usepackage{placeins}
\usepackage{enumerate}
\usepackage[usenames,dvipsnames]{xcolor}
\usepackage[colorlinks]{hyperref}
\usepackage{setspace}

\usepackage{tikz}
\usetikzlibrary{calc}
\usetikzlibrary{shapes.geometric, arrows}
\usetikzlibrary{arrows.meta}
\tikzset{>={Latex[width=1mm,length=1mm]}}

\tikzstyle{process} = [rectangle, minimum width=2.5cm, minimum height=1cm, text
centered, text width = 3cm, draw=black]
\tikzstyle{decision} = [diamond, minimum width=2.5cm, minimum height=1cm,
aspect=2,inner sep=-0.5ex,text centered, text width = 2.5cm, draw=black]
\tikzstyle{arrow} = [thick,->,>=stealth]
\tikzstyle{line} = [draw, -latex']

\biboptions{sort&compress}
\newcommand{\revA}[1]{{\color{black} #1}}
\newcommand{\revB}[1]{{\color{black} #1}}

\journal{Journal of Computational Physics}

\newcommand\vecu{\mathbf{u}}
\newcommand\imp{^{(n+1)}} 
\newcommand\rhs{^{(n)}} 
\newcommand\pts{^{(t-\Delta t)}} 
\newcommand\N{\mathcal{N}}
\newcommand\D{\mathcal{D}}

\begin{document}

\begin{frontmatter}

\title{Breaching the capillary time-step constraint using a coupled VOF method with implicit surface tension}

\author{Fabian Denner\corref{cor1}}
\author{Fabien Evrard}
\author{Berend van Wachem}

\address{Chair of Mechanical Process Engineering, Otto-von-Guericke-Universit\"{a}t Magdeburg,\\ Universit\"atsplatz 2, 39106 Magdeburg, Germany}

\cortext[cor1]{fabian.denner@ovgu.de}

\begin{abstract}
The capillary time-step constraint is the dominant limitation on the applicable time-step in many simulations of interfacial flows with surface tension and, consequently, governs the execution time of these simulations. We propose a fully-coupled pressure-based algorithm based on an algebraic Volume-of-Fluid (VOF) method in conjunction with an implicit linearised surface tension treatment that can breach the capillary time-step constraint. The advection of the interface is solved together with the momentum and continuity equations of the interfacial flow in a single system of linearised equations, providing an implicit coupling between pressure, velocity and the VOF colour function used to distinguish the interacting fluids. Surface tension is treated with an implicit formulation of the Continuum Surface Force (CSF) model, whereby both the interface curvature and the gradient of the colour function are treated implicitly with respect to the colour function. 
The presented results demonstrate that a time-step larger than the capillary time-step can be applied with this new numerical framework, as long as other relevant time-step restrictions are satisfied, including a time-step restriction associated with surface tension, density as well as viscosity.
\end{abstract}

\begin{keyword}
Capillary time-step constraint \sep Surface tension \sep Coupled algorithm \sep Volume-of-Fluid method \\~\\
\textcopyright~2022. This manuscript version is made available under the CC-BY-NC-ND 4.0 license. \href{http://creativecommons.org/licenses/by-nc-nd/4.0/}{http://creativecommons.org/licenses/by-nc-nd/4.0/}
\end{keyword}

\end{frontmatter}

\section{Introduction}
The temporal resolution of the propagation of the smallest capillary waves discretely resolved in space on a fluid interface presents the dominant time-step restriction for the majority of simulations of interfacial flows with surface tension \citep{Popinet2018}. It has long been speculated that an implicit formulation of surface tension must be able to eliminate, or at least mitigate, this time-step constraint \citep{Brackbill1992, Kothe1998, Hysing2006,Popinet2009,Sussman2009,Popinet2018}. To date, however, no numerical algorithm based on an interface capturing method, such as a Volume-of-Fluid (VOF), level-set or phase-field method, has been reported in the literature that can breach the capillary time-step constraint, while still representing the governing physics faithfully and without adding unphysical terms to the governing equations.

The capillary time-step constraint arises from the phase velocity $c$ of capillary waves, given as \citep{Lamb1932}
\begin{equation}
    c = \sqrt{\frac{2 \pi \sigma}{(\rho_\text{a}+\rho_\text{b}) \lambda}},
    \label{eq:c}
\end{equation}
where $\sigma$ is the surface tension coefficient, $\lambda$ is the wavelength, and $\rho_\text{a}$ and $\rho_\text{b}$ are the densities of the interacting fluids. Because the phase velocity is inversely proportional to $\sqrt{\lambda}$, shorter capillary waves with an increasing phase velocity are spatially resolved as the computational mesh is refined. 
\citet{Brackbill1992} were the first to recognise this time-step restriction. 
\revB{Considering that the shortest unambiguously resolved capillary waves have a wavenumber of $k_\sigma = \pi/\Delta x$ and, consequently, a wavelength of $\lambda_\sigma=2\pi/k_\sigma=2\Delta x$, where $\Delta x$ is the mesh spacing, \citet{Brackbill1992} proposed the capillary time-step constraint as}
\begin{equation}
    \Delta t_\sigma = \frac{\Delta x}{2 c} = \sqrt{\frac{\rho_\text{a}+\rho_\text{b}}{4 \pi \sigma} \Delta x^3}.
    \label{eq:tsigma_Brackbill}
\end{equation}
The capillary time-step constraint is, therefore, a Courant–Friedrichs–Lewy (CFL) condition \citep{Courant1928} associated with the phase velocity of capillary waves, $\Delta t_\sigma \propto \Delta x/c$, with the factor 2 in Eq.~\eqref{eq:tsigma_Brackbill} accounting for the case in which two oppositely propagating waves enter the same mesh cell simultaneously. As part of our previous work \citep{Denner2015}, we revisited the origin of the capillary time-step constraint using both numerical and signal-processing arguments, and arrived at a similar but slightly different formulation of the capillary time-step constraint than \citet{Brackbill1992}, given as\footnote{The effective phase velocity in the case where two oppositely propagating waves enter the same mesh cell simultaneously increases by factor $\sqrt{2}$, a result directly following from Eq.~\eqref{eq:c}, rather than factor 2 \citep{Denner2015}.}
\begin{equation}
    \Delta t_\sigma = \frac{\Delta x}{\sqrt{2} \, c} =\sqrt{\frac{\rho_\text{a}+\rho_\text{b}}{2 \pi \sigma} \Delta x^3}.
    \label{eq:tsigma_Denner}
\end{equation}
 We also demonstrated that, if the velocity of the flow at the interface has a magnitude similar to the phase velocity $c$, the Doppler shift associated with capillary waves propagating along a moving interface ought to be taken into account, and provided the first numerical results that clearly delineate the capillary time-step constraint.

Contrary to the traditional CFL condition, $\Delta t \leq \Delta x / |\vecu|$ \citep{Courant1928}, which arises from the flow velocity $\vecu$ and is proportional to the mesh spacing $\Delta x$, the capillary time-step constraint is proportional to $\Delta x^{3/2}$, since $c \propto \Delta x^{-1/2}$. As a consequence, the capillary time-step constraint dominates the maximum applicable time-step for interfacial flows at small lengthscales ({\it e.g.}~microfluidics), applications with quasi-static heat and mass transfer ({\it e.g.}~the evaporation of a sessile drop) and, in general, simulations with high spatial resolution, as it is now routinely afforded by adaptive mesh refinement algorithms in conjunction with modern high-performance computing resources.
Even at very small scales, such as in microfluidic applications, viscous contributions with a physical origin are typically not able to mitigate the capillary time-step constraint \citep{Popinet2018, Denner2015}.

Already in their original proposition of the capillary time-step constraint, \citet{Brackbill1992} hypothesised that an implicit treatment of surface tension should allow to breach or even eliminate the capillary time-step constraint. 
\citet{Hou1994a} presented a boundary integral method with an implicit surface tension treatment for irrotational incompressible flows in two dimensions that is evidently free of the capillary time-step constraint, demonstrating the necessity of an implicit surface tension treatment in eliminating the capillary time-step constraint.
Following the work of \citet{Bansch2001}, \citet{Hysing2006} proposed a semi-implicit surface tension treatment for a two-dimensional finite-element method, which, in essence, incorporates the interface position at the new time instance implicitly. 
\citet{Raessi2009} translated this surface tension treatment subsequently to finite-volume methods. \citet{Hysing2006} and \citet{Raessi2009} reported stable results for time-steps exceeding the capillary time-step constraint, although the solution was not stable for arbitrarily large time-steps (assuming other relevant time-step restrictions, {\it e.g.}~the CFL condition, are satisfied). 
We previously proposed an algorithm in which the continuity, momentum and VOF advection equations are implicitly coupled and solved as a single system of equations \citep{Denner2015}, treating the surface tension term semi-implicit with respect to the VOF colour function. However, this method does not allow to breach the capillary time-step constraint, which led us to conclude \revB{(incorrectly)} that an implicit formulation of the source term representing surface tension cannot eliminate the capillary time-step constraint using interface capturing methods \citep{Denner2015}. A fully-Lagrangian method for incompressible free-surface flows was proposed by \citet{Zheng2015}, based on a Marker-and-Cell (MAC) method in conjunction with a Lagrangian fluid mesh in the vicinity of the fluid interface, which are coupled implicitly and ensure an exact balance between surface tension and pressure. To date, this is the only method for three-dimensional interfacial flows that has been demonstrated to be practically free of the capillary time-step constraint, whereas a numerical framework based on an interface capturing method with comparable capabilities has yet to be published.

In this article, we propose a numerical framework for interfacial flows based on a VOF method that is able to breach the capillary time-step constraint. We achieve this by extending a fully-coupled pressure-based algorithm \citep{Denner2020} with an algebraic VOF method that is implicitly coupled to the governing flow equations, enabling an implicit treatment of surface tension based on the Continuum Surface Force (CSF) model \citep{Brackbill1992}. The presented results demonstrate that a time-step larger than the capillary time-step constraint can be applied with this new numerical framework, as long as other relevant time-step restrictions ({\em e.g.}~the CFL condition) are satisfied. However, a new time-step constraint arises that depends on surface tension, density and viscosity, but which is less restrictive than the classical capillary time-step constraint.

\section{Mathematical model}
In this study, we consider interfacial flows of two immiscible and incompressible Newtonian fluids. Such an interfacial flow is governed by the continuity equation 
\begin{equation}
    \frac{\partial u_i}{\partial x_i} = 0 ,
    \label{eq:continuity}  
\end{equation}
with $\mathbf{x}$ the spatial coordinate and $\vecu$ the velocity vector, and the momentum equations
\begin{equation}
    \rho \left( \frac{\partial u_j}{\partial t} + \frac{\partial u_i u_j}{\partial x_i} \right) = - \frac{\partial p}{\partial x_j} + \frac{\partial \tau_{ji}}{\partial x_i} + S_j ,
    \label{eq:momentum}
\end{equation}
where $\rho$ is the density, $t$ denotes time, $p$ is the pressure and $\mathbf{S}$ is the source term representing surface tension.
The stress tensor $\boldsymbol{\tau}$ is given as
\begin{equation}
    \tau_{ji} = \mu \left(\frac{\partial u_j}{\partial x_i} + \frac{\partial u_i}{\partial x_j} \right) ,
    \label{eq:stresstensor}
\end{equation}
where $\mu$ is the dynamic viscosity.

A VOF method \citep{Hirt1981} is adopted to model the transport and interaction of two immiscible fluids. The two fluids are distinguished by the indicator function $\zeta$, which is defined as
\begin{equation}
    \zeta(\mathbf{x}) = 
    \begin{cases}
        0, &\text{if} \ \mathbf{x} \in \Omega_\text{a}\\  
        1, &\text{if} \ \mathbf{x} \in \Omega_\text{b}  
    \end{cases}
\end{equation}
where $\Omega = \Omega_\text{a} \cup \Omega_\text{b}$ is the computational domain, with $\Omega_\text{a}$ and $\Omega_\text{b}$ the subdomains occupied by fluids ``a'' and ``b'', respectively.
The indicator function $\zeta$ is advected by the underlying flow
\begin{equation}
   \frac{\text{D}\zeta}{\text{D}t} = \frac{\partial \zeta}{\partial t} +  \frac{\partial u_i \zeta}{\partial x_i} - \zeta \frac{\partial u_i}{\partial x_i} = 0.
    \label{eq:vofadvection_prod}
\end{equation}
The fluid properties are defined over the entire computational domain using the indicator function, {\em e.g.}~for density $\rho = (1-\zeta) \rho_a + \zeta \rho_b$.
However, to focus the discussion  on the discretisation and influence of surface tension, both interacting fluids are assumed to have the same density, $\rho=\rho_\text{a} = \rho_\text{b}$, and viscosity, $\mu=\mu_\text{a} = \mu_\text{b}$, throughout this study.

\section{Numerical framework}
\label{sec:numerics}

\begin{figure}[t]
\begin{center}
\begin{small}
    \begin{tikzpicture}
    \node (pro0) [process] {Update \mbox{previous} time-levels: $\chi^{(t-2\Delta t)}\leftarrow \chi^{(t-\Delta t)}$ $\chi^{(t-\Delta t)}\leftarrow \chi^{(n)\phantom{...)}}$ $\vartheta_f^{(t-\Delta t)} \leftarrow \vartheta_f^{(n)\phantom{....}}$};
    \node (pro1a) [process, below of=pro0, yshift=-1cm] {\mbox{Gather coefficients} \mbox{and assemble}  \mbox{$\boldsymbol{\mathcal{A}}$ and $\mathbf{b}$}}; 
    \node (pro1b) [process, below of=pro1a, yshift=-0.5cm] {Solve $\boldsymbol{\mathcal{A}} \cdot \boldsymbol{\phi} = \mathbf{b}$}; 
    \node (pro2) [process, below of=pro1b, yshift=-0.5cm] {Update \mbox{deferred} quantities: $\chi^{(n)}
    \leftarrow \chi^{(n+1)}$};
    \node (pro3) [process, below of=pro2, yshift=-0.5cm] {Compute $\kappa^{(n)}$ and $\vartheta_f^{(n)}$};
    \node (dec1) [decision, below of=pro3, yshift=-0.7cm]
    {Conservation satisfied?}; 
    \draw [arrow] (pro0) -- (pro1a);
    \draw [arrow] (pro1a) -- (pro1b);
    \draw [arrow] (pro1b) -- (pro2);
    \draw [arrow] (pro2) -- (pro3);
    \draw [arrow] (pro3) -- (dec1);
    \draw [arrow] (dec1) --+(-2.7cm,0) |- (pro1a);
    \draw [arrow] (dec1) --+(+2.7cm,0) |- (pro0);
    \node at (-2.1,-8) {no};
    \node at (2.1,-8) {yes};
    \node [rotate=90] at (-2.95,-5) {$n=n+1$};
    \node [rotate=90] at (2.95,-5) {$t=t+\Delta t$};
    \end{tikzpicture} 
\end{small}
\caption{Flow chart of the solution procedure of the discretised and linearised system of governing equations, where $n$ is the nonlinear iteration counter, $\chi \in \{p,u,v,w,\psi \}$ are the solution variables, $\kappa$ is the interface curvature (see Section \ref{sec:surfacetension}) and $\vartheta_f$ is the advecting velocity through mesh face $f$ (see Section \ref{sec:mwi}). The coefficient matrix $\boldsymbol{\mathcal{A}}$ holds all coefficients for the implicitly sought solution variables $\chi^{(n+1)}$ of the discretised governing equations (see Section \ref{sec:solution}) and $\boldsymbol{\phi}$ is the solution vector. The right-hand side vector $\mathbf{b}$ holds the deferred contributions of the previous iteration ($\chi^{(n)}$, $\kappa^{(n)}$, $\vartheta_f^{(n)}$) and the contributions of the previous time-levels ($\chi^{(t-\Delta t)}$, $\chi^{(t-2\Delta t)}$, $\vartheta_f^{(t-\Delta t)}$).}
\label{fig:flowchart}
\end{center}
\end{figure}
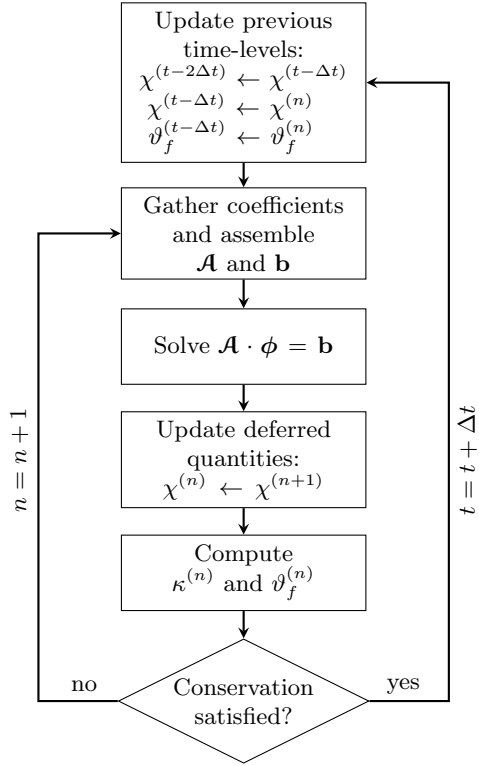

The proposed numerical framework builds upon a class of fully-coupled pressure-based algorithms for single-phase \citep{Denner2018c, Denner2020} and interfacial flows \citep{Denner2014a, Denner2015}, with the aim of treating all discretised and linearised governing equations implicitly in a single linear system of governing equations, $\boldsymbol{\mathcal{A}} \cdot \boldsymbol{\phi} = \mathbf{b}$.
This linear system of governing equations is solved simultaneously for the pressure $p$, the velocity $\mathbf{u} \equiv (u~v~w)^\text{T}$ and the discrete colour function $\psi$, using the Block-Jacobi pre-conditioner and the BiCGSTAB solver of the software library PETSc \citep{petsc-user-ref,petsc-web-page}. The nonlinear nature of the governing equations is accounted for by means of an inexact Newton method \citep{Dembo1982}, whereby the deferred terms resulting from the applied linearisation procedure are updated iteratively, until the nonlinear system of governing equations satisfies predefined conservation criteria, as illustrated in Figure \ref{fig:flowchart}. The discretisation of the governing equations is based on an established second-order finite-volume method and utilises a collocated variable arrangement \citep{Denner2020}, with the fluxes through the mesh faces computed by a momentum-weighted interpolation \citep{Bartholomew2018}. The proposed numerical framework is not inherently limited to VOF methods, but may similarly be used in conjunction with level-set or phase-field methods.

\subsection{Discretised governing equations}

In the following we assume an equidistant Cartesian mesh with mesh spacing $\Delta x$.
The continuity equation \eqref{eq:continuity} for each computational cell $P$ is readily discretised as
\begin{equation}
\sum_f F_f\imp = 0, \label{eq:continuity_disc}
\end{equation}
where subscript $f$ denotes faces adjacent to mesh cell $P$, $F_f = \vartheta_f A_f$ is the flux through face $f$, $A_f$ is the area of face $f$ and $\vartheta_f$ is the advecting velocity (outward pointing with respect to cell $P$) obtained from a momentum-weighted interpolation, as further detailed in Section \ref{sec:mwi}. The superscript $(n+1)$ denotes implicitly solved quantities and the superscript $(n)$ denotes deferred quantities, where $n$ is the nonlinear iteration counter.
Applying the second-order backward Euler scheme to discretise the transient term \citep{Ferziger2020}, the discretised momentum equations \eqref{eq:momentum} follow as
\begin{equation}
    \begin{split}
        & \rho \left[\frac{3 u_{j,P}\imp - 4 u_{j,P}\pts + u_{j,P}^{(t-2\Delta t)}}{2\Delta t}  V_P +  \sum_f \left( \tilde{u}_{j,f}\imp F_f\rhs + \tilde{u}_{j,f}\rhs F_f\imp - \tilde{u}_{j,f}\rhs F_f\rhs \right)\right] \\ &= - \sum_f \overline{p}_f\imp n_{j,f} A_f + \mu \sum_f \left( \frac{u_{j,Q}\imp - u_{j,P}\imp}{\Delta x} \,  + \left. \overline{\frac{\partial {u}_{i}}{\partial x_j}}\right|_f\imp n_{i,f}\right)  A_f + {S}_{j,P}\imp V_P ,
    \end{split}
    \label{eq:momentum_disc}
\end{equation}
where $\tilde{\square}$ denotes flux-limited values at face $f$ (see below), $\overline{\square}_f=(\square_P+\square_Q)/2$ denotes a linear interpolation of the cell-centred values to face $f$, with $Q$ the neighbour cell of $P$ adjacent to face $f$, $\mathbf{n}_f$ is the normal vector of face $f$ (outward pointing with respect to cell $P$) and, as further described in Section \ref{sec:surfacetension}, $\mathbf{S}$ is the source term representing surface tension. A Newton linearisation is applied to the advection term of the momentum equations \citep{Denner2018c}, to enable an implicit treatment of both the velocity $\tilde{\mathbf{u}}_f$ and the flux $F_f$. 

The two interacting fluids are represented discretely by the colour function $\psi$, 
\begin{equation}
    \psi_P = \frac{1}{V_P} \iiint_{V_P} \zeta \, \text{d}V,
    \label{eq:discretecolour}
\end{equation}
which is advected based on Eq.~\eqref{eq:vofadvection_prod} by
\begin{equation}
    \begin{split}
    \frac{3 \psi_P\imp - 4 \psi_P\pts + \psi_P^{(t-2\Delta t)}}{2\Delta t}  V_P &+ \sum_f \left( \tilde{\psi}_f\imp F_f\rhs + \tilde{\psi}_f\rhs F_f\imp - \tilde{\psi}_f\rhs F_f\rhs \right) \\ 
    &- \left(\psi_P\imp \sum_f F_f\rhs + \psi_P\rhs \sum_f F_f\imp - \psi_P\rhs \sum_f F_f\rhs \right)
    = 0.
    \end{split}
    \label{eq:vof_disc}
\end{equation}
In analogy to the discretisation of the momentum equations \eqref{eq:momentum_disc}, the transient term is discretised with the second-order backward Euler scheme and both spatial terms are linearised with a Newton linearisation. 

The flux-limited face values are given as
\begin{equation}
\tilde{\psi}_f\imp = 
\begin{cases}
    \psi_U\imp + \xi^{(\psi)}_f \left(\psi_D\imp-\psi_U\imp\right), ~~&\text{if}~\left|\psi_U\rhs-\psi_D\rhs\right| > \varepsilon_\text{tol}\\ 
\psi_U\imp,~~&\text{else}\\ 
\end{cases} 
\label{eq:differencing_colour}
\end{equation}
and
\begin{equation}
    \tilde{\vecu}_f\imp = 
    \begin{cases}
    \vecu_U\imp + \xi^{(\psi)}_f \left(\vecu_D\imp-\vecu_U\imp\right), ~~&\text{if}~\left|\psi_U\rhs -\psi_D\rhs\right| > \varepsilon_\text{tol}\\ 
    \dfrac{\vecu_P\imp + \vecu_Q\imp}{2},~~&\text{else}\\ 
    \end{cases}
\end{equation}
where $\xi^{(\psi)}_f$ is the flux limiter of the colour function, $\varepsilon_\text{tol} = 10^{-6}$ is a predefined tolerance, and subscripts $U$ and $D$ denote the upwind and downwind cells of face $f$, respectively. For the purpose of this study, the CICSAM scheme \citep{Ubbink1999} is used to compute the flux limiter $\xi^{(\psi)}_f$, but any other advection scheme may also be applied. Note that the advection scheme applied to the colour function in Eq.~\eqref{eq:differencing_colour} away from the interface, where $\psi = \text{const.}$, is irrelevant and choosing upwind differencing is here a matter of convenience.

\subsection{Surface tension}
\label{sec:surfacetension}
Surface tension is modelled using the CSF model \citep{Brackbill1992}, with which the source term $\mathbf{S}$ representing surface tension is given as
\begin{equation}
    \mathbf{S} = \sigma \, \kappa \, \nabla \psi,
    \label{eq:csf}
\end{equation}
where $\sigma$ is the surface tension coefficient and $\kappa$ is the interface curvature. 
\revA{The gradient of the colour function is discretised using the Gauss theorem, analogous to the discretisation of the pressure gradient in the momentum equations \eqref{eq:momentum_disc}, thus it is given as
\begin{equation}
     \nabla \psi_P \approx \frac{1}{V_P} \sum_f \overline{\psi}_f \mathbf{n}_f A_f, \label{eq:colourgrad}
 \end{equation}
which facilitates a force-balanced discretisation \citep{Denner2014a}.}
The source term $\mathbf{S}$ is treated implicitly by applying a Newton linearisation of the interface curvature and the gradient of the colour function to yield
\revA{
    \begin{equation}
        \mathbf{S}_P\imp \approx \sigma \kappa_P\rhs  \nabla \psi_P\imp +  \sigma \kappa_P\imp \nabla \psi_P\rhs -  \sigma \kappa_P\rhs \nabla \psi_P\rhs, \label{eq:csf_imp0}
     \end{equation}
which, by inserting Eq.~\eqref{eq:colourgrad}, becomes}
\begin{equation}
    \mathbf{S}_P\imp \approx \frac{\sigma}{V_P} \left(\kappa_P\rhs \sum_f \overline{\psi}_f\imp \mathbf{n}_{f} A_f + \kappa_P\imp  \sum_f \overline{\psi}_f\rhs \mathbf{n}_{f} A_f -  \kappa_P\rhs  \sum_f \overline{\psi}_f\rhs \mathbf{n}_{f} A_f  \right). \label{eq:csf_imp}
\end{equation}

Various approaches have been proposed for estimating the interface curvature from the discrete colour function, {\it e.g.}~\citep{Brackbill1992, Cummins2005, Raessi2007, Popinet2009, Denner2014a, Owkes2014a, Evrard2017, Evrard2020}. In this work, we employ the height-function (HF) method \cite{Evrard2020} both for its superior accuracy and ease of linearisation. In the HF method, when the $z$-component of the interface normal is dominant, curvature reads as
\begin{equation}
    \kappa = \frac{-H_{xx}\left(1+H_y^2\right) - H_{yy}\left(1+H_x^2\right)+2H_xH_yH_{xy}}{\left(H_x^2+H_y^2+1\right)^{3/2}}, 
\end{equation}
where $H_{\{x,y,xx,yy,xy\}}$ are the first and second partial derivatives of the ``heights of fluid'' computed along the $z$-direction\footnote{In case the $x$- or $y$-component of the normal is dominant, the indices of the partial derivatives are simply permuted.}. On a Cartesian mesh, a height of fluid can be trivially obtained by summing the colour function in a column of computational cells. The partial derivatives of the heights are calculated using central differences, therefore they can be expressed as linear combinations of the discrete colour function values in the set of cells from which the heights are computed. Consider, for instance, an interfacial computational cell $P$, $\psi_P \in \left]0,1\right[$, within which the $z$-component of the interface normal is dominant. One can construct a stencil $\mathcal{S}(P)$ centred around $P$, containing at most $3\times3\times N_H$ cells, with $N_H$ an odd number typically chosen between $5$ and $9$. For the portion of interface in cell $P$, the first partial derivative of the heights along the $x$-direction is approximated as
\begin{equation}
    H_x = \sum_{N\in\mathcal{S}(P)} \beta_{x,N} \, \psi_N ,
\end{equation}
where $\beta_{x,N}$ are coefficients arising from the application of central differences to the heights, themselves calculated as the sum of the discrete colour function values in the columns of $N_H$ computational cells of $\mathcal{S}(P)$. The other partial derivatives of the heights are expressed similarly. 

In the cell $P$, applying a Newton linearisation on curvature yields
\begin{equation}
    \kappa\imp_P = \kappa\rhs_P + \sum_{N\in\mathcal{S}(P)} \left[\left(\psi\imp_N-\psi\rhs_N\right) \left(\frac{\partial \kappa_P}{\partial \psi_N}\right)\rhs \right].
    \end{equation}
Introducing the quantities
\begin{align}
    \N_P & = -H_{xx}\left(1+H_y^2\right) - H_{yy}\left(1+H_x^2\right)+2H_xH_yH_{xy} \\
    \D_P & = H_x^2+H_y^2+1
\end{align}
the linearisation of curvature can be reformulated as
\begin{equation}
\kappa\imp_P = \kappa\rhs_P + \frac{1}{\D_P\rhs} \sum_{N\in\mathcal{S}(P)} \left[\left(\psi\imp_N-\psi\rhs_N\right) \left(\left(\frac{\partial \N_P}{\partial \psi_N}\right)\rhs - \frac{3}{2} \kappa\rhs_P \sqrt{\D_P\rhs} \left(\frac{\partial \D_P}{\partial \psi_N}\right)\rhs\right) \right]. \label{eq:Klinear}
\end{equation}
The partial derivatives on the right-hand side of Eq.~\eqref{eq:Klinear} read as
\begin{align}
    \frac{\partial \N_P}{\partial \psi_N} & = -\left(1+H_{y}^2\right)\beta_{xx,N} -\left(1+H_{x}^2\right)\beta_{yy,N} \nonumber \\ & \quad \quad \quad \quad \quad 
    +2\left(H_yH_{xy}-H_{x}H_{yy}\right)\beta_{x,N}
    +2\left(H_xH_{xy}-H_yH_{xx}\right)\beta_{y,N}
    +2H_xH_y\beta_{xy,N} ,
    \\
    \frac{\partial \D_P}{\partial \psi_N} & = 2 H_{x} \beta_{x,N} + 2 H_{y} \beta_{y,N}.
\end{align}
Note that for two-dimensional flow in the plane $(x,z)$, the previous expressions reduce to 
\begin{align}
    \N_P & = -H_{xx} , &  \D_P & = H_x^2+1,    \\
    \frac{\partial \N_P}{\partial \psi_N} & = -\beta_{xx,N} , &   \frac{\partial \D_P}{\partial \psi_N} & = 2 H_{x} \beta_{x,N}.
\end{align}

\subsection{Momentum-weighted interpolation}
\label{sec:mwi}

The advecting velocity $\vartheta_f=\vecu_f \cdot \mathbf{n}_f$ of the flux $F_f = \vartheta_f A_f$ through face $f$ is discretised using a momentum-weighted interpolation (MWI) \citep{Rhie1983}, which provides a direct coupling of pressure and velocity that eliminates pressure-velocity decoupling as a result of the applied collocated variable arrangement. 
Including the source term representing surface tension, the advecting velocity is defined as \citep{Denner2014a, Bartholomew2018}
\begin{equation}
    \vartheta_f = \overline{\vecu}_{f} \cdot \mathbf{n}_{f} - \hat{d}_f \left(  \nabla p_f -   \overline{\nabla p}_f \right) \cdot \mathbf{n}_{f}  + \hat{d}_f \left(  \mathbf{S}_f -  \overline{\mathbf{S}}_f \right) \cdot \mathbf{n}_{f} + \hat{d}_f \frac{\rho}{\Delta t} \left( \vartheta_f\pts - \overline{\vecu}_{f}\pts \cdot \mathbf{n}_{f} \right).
    \label{eq:mwi}
\end{equation}
The coefficient $\hat{d}_f$, derived in detail in \citep{Bartholomew2018}, represents the weighting factor of the MWI correction terms and, in the context of a fully-coupled algorithm, the strength of the implicit coupling provided by the MWI. 
The pressure gradient at the face is discretised as
\begin{equation}
    \nabla p_f \cdot \mathbf{n}_f \approx \frac{p_Q-p_P}{\Delta x}
\end{equation}
and, analogous to the pressure term in the momentum equations \eqref{eq:momentum_disc}, the cell-centred pressure gradient is discretised using the Gauss theorem as
\begin{equation}
    \nabla p_P \approx \frac{1}{V_P} \sum_f \overline{p}_f \mathbf{n}_f A_f.
\end{equation}
Together, the pressure terms in Eq.~\eqref{eq:mwi}$, (\nabla p_f - \overline{\nabla p}_f)$, then act as a low-pass filter with respect to pressure \citep{Bartholomew2018}.
Similarly, the surface tension terms are discretised as \citep{Denner2014a}
\begin{equation}
    \mathbf{S}_f \cdot \mathbf{n}_f \approx \sigma \overline{\kappa}_f \frac{\psi_Q-\psi_P}{\Delta x}
\end{equation}
and
\begin{equation}
    \mathbf{S}_P \approx \sigma \kappa_P \nabla \psi_P = \frac{\sigma \kappa_P}{V_P} \sum_f \overline{\psi}_f \mathbf{n}_f A_f.
\end{equation}

Applying the discretisation described above, along with a Newton linearisation of the surface tension terms analogous to the source term representing surface tension in the momentum equations, the advecting velocity is proposed as
\begin{equation}
\begin{split}
    \vartheta_f\imp = \overline{\vecu}_{f}\imp \cdot \mathbf{n}_{f} &-~\hat{d}_f \left[\frac{p_Q\imp-p_P\imp}{\Delta x} - \frac{1}{2} \left( \nabla p_P\imp +  \nabla p_Q\imp \right) \cdot \mathbf{n}_{f} \right] \\ 
    &+~\hat{d}_f \sigma \left[\overline{\kappa}_f\rhs \frac{\psi_Q\imp-\psi_P\imp}{\Delta x} - \frac{1}{2} \left(\kappa_P\rhs \nabla \psi_P\imp + \kappa_Q\rhs \nabla \psi_Q\imp \right)  \cdot \mathbf{n}_{f} \right]\\ 
    &+~\hat{d}_f \sigma \left[\overline{\kappa}_f\imp \frac{\psi_Q\rhs-\psi_P\rhs}{\Delta x} - \frac{1}{2} \left(\kappa_P\imp \nabla \psi_P\rhs + \kappa_Q\imp \nabla \psi_Q\rhs \right)  \cdot \mathbf{n}_{f} \right]\\ 
    &-~\hat{d}_f \sigma \left[\overline{\kappa}_f\rhs \frac{\psi_Q\rhs-\psi_P\rhs}{\Delta x} - \frac{1}{2} \left(\kappa_P\rhs \nabla \psi_P\rhs + \kappa_Q\rhs \nabla \psi_Q\rhs \right)  \cdot \mathbf{n}_{f} \right] \\ 
    &+~\hat{d}_f \frac{\rho}{\Delta t} \left( \vartheta_f\pts - \overline{\vecu}_{f}\pts \cdot \mathbf{n}_{f} \right).
\end{split}
\label{eq:mwi_implicit}
\end{equation}
With this discrete formulation, each term of the current time-level in the advecting velocity, Eq.~\eqref{eq:mwi}, makes an implicit contribution to the solution variables $\chi \in \{p,u,v,w,\psi\}$.

\subsection{Solution procedure}
\label{sec:solution}
The discretised governing equations \eqref{eq:continuity_disc}, \eqref{eq:momentum_disc} and \eqref{eq:vof_disc} are solved simultaneously in a single linear system of discretised equations, given for a three-dimensional mesh with $N$ cells as 
\begin{equation}
    \begin{pmatrix}
        \boldsymbol{\mathcal{A}}^{p}_\text{cont.} 
        & \boldsymbol{\mathcal{A}}^u_\text{cont.} 
        & \boldsymbol{\mathcal{A}}^v_\text{cont.} 
        & \boldsymbol{\mathcal{A}}^w_\text{cont.}
        & \boldsymbol{\mathcal{A}}^\psi_\text{cont.} \\
        \boldsymbol{\mathcal{A}}^{p}_\text{$x$-mom.} 
        & \boldsymbol{\mathcal{A}}^u_\text{$x$-mom.} 
        & \boldsymbol{\mathcal{A}}^v_\text{$x$-mom.} 
        & \boldsymbol{\mathcal{A}}^w_\text{$x$-mom.}
        & \boldsymbol{\mathcal{A}}^\psi_\text{$x$-mom.} \\
        \boldsymbol{\mathcal{A}}^{p}_\text{$y$-mom.} 
        & \boldsymbol{\mathcal{A}}^u_\text{$y$-mom.} 
        & \boldsymbol{\mathcal{A}}^v_\text{$y$-mom.} 
        & \boldsymbol{\mathcal{A}}^w_\text{$y$-mom.}
        & \boldsymbol{\mathcal{A}}^\psi_\text{$y$-mom.} \\
        \boldsymbol{\mathcal{A}}^{p}_\text{$z$-mom.} 
        & \boldsymbol{\mathcal{A}}^u_\text{$z$-mom.} 
        & \boldsymbol{\mathcal{A}}^v_\text{$z$-mom.} 
        & \boldsymbol{\mathcal{A}}^w_\text{$z$-mom.}
        & \boldsymbol{\mathcal{A}}^\psi_\text{$z$-mom.} \\
        \boldsymbol{\mathcal{A}}^{p}_\text{\sc vof} 
        & \boldsymbol{\mathcal{A}}^u_\text{\sc vof} 
        & \boldsymbol{\mathcal{A}}^v_\text{\sc vof} 
        & \boldsymbol{\mathcal{A}}^w_\text{\sc vof}
        & \boldsymbol{\mathcal{A}}^\psi_\text{\sc vof} \\
    \end{pmatrix}
    \cdot
    \begin{pmatrix}
        \boldsymbol{\phi}^p \\ 
        \boldsymbol{\phi}^u \\ 
        \boldsymbol{\phi}^v \\ 
        \boldsymbol{\phi}^w \\
        \boldsymbol{\phi}^\psi
    \end{pmatrix}
    =
    \begin{pmatrix}
        \mathbf{b}_\text{cont.} \\ 
        \mathbf{b}_\text{$x$-mom.} \\ 
        \mathbf{b}_\text{$y$-mom.} \\ 
        \mathbf{b}_\text{$z$-mom.} \\
        \mathbf{b}_\text{\sc vof}
    \end{pmatrix}, 
    \label{eq:eqsysfull}
\end{equation}
for the solution variables pressure $p$, velocity $\vecu \equiv (u~v~w)^\text{T}$ and colour function $\psi$. In Eq.~\eqref{eq:eqsysfull}, $\boldsymbol{\mathcal{A}}^\chi_\text{eq.}$ denotes the coefficient submatrix of size $N \times N$ of the continuity equation (eq.~= cont.), the momentum equations associated with the three Cartesian coordinate axes (eq.~= $x$-mom., eq.~= $y$-mom., eq.~= $z$-mom.) and the VOF advection equation (eq.~= {\sc vof}) for the respective solution variable $\chi \in \{p,u,v,w,\psi\}$. The solution subvectors of length $N$ for solution variable $\chi$ are denoted as $\boldsymbol{\phi}^\chi$ and the right-hand side subvectors of length $N$ of the five discretised governing equations, which contain all deferred contributions and contributions from previous time-levels, are denoted as $\mathbf{b}_\text{eq.}$. This fully-coupled system of equations~\eqref{eq:eqsysfull} is solved in an iterative fashion using the Block-Jacobi pre-conditioner and the BiCGSTAB solver of the software library PETSc \citep{petsc-user-ref,petsc-web-page}, as illustrated in Figure \ref{fig:flowchart}.

The Newton linearisation of the advection terms in both the momentum equations and the VOF advection equation yields an implicit contribution of the fluxes, which in turn introduces an implicit pressure, velocity and colour function dependency in all governing equations. The implicit formulation of the fluxes by the MWI presented in Eq.~\eqref{eq:mwi_implicit} is, thus, the primary coupling term of the discretised governing equations. Notably, the implicit pressure and velocity dependency in the VOF advection is a novel building block for solving interfacial flows. Both the colour function gradient and the curvature are treated implicitly with respect to the colour function, yielding an implicit CSF model for surface tension.

\section{Differences to previously proposed methods}
\label{sec:comparisonprevious}

Previous work aimed at breaching the capillary time-step constraint in the context of interface capturing methods, as already mentioned in the introduction of this article, has focused on incorporating the interface position at the new time instance implicitly in the momentum equations \citep{Hysing2006, Raessi2009} and on a coupled solution algorithm in which all governing equations are solved simultaneously \citep{Denner2015}.

\citet{Hysing2006} proposed a semi-implicit surface tension treatment in the context of a finite-element method, whereby the interface position at the new time instance is incorporate implicitly in the source term representing surface tension.
Using the CSF method to model surface tension in a finite-volume discretisation, the source term representing surface tension, accounting for the interface position at the new time instance implicitly, is given as \citep{Raessi2009}
\begin{equation}
    \mathbf{S}_P^{(n+1)} \approx \sigma \, \kappa_P\rhs \, \nabla \psi_P\rhs + \sigma \, \Delta t \, |\nabla \psi_P\rhs| \, \Delta_\text{s} \vecu_P\imp,
    \label{eq:surfacetensionRaessi}
\end{equation}
where $\Delta_\text{s}$ is the Laplace-Beltrami operator with respect to the interface. This formulation is convenient since the implementation in existing implicit numerical frameworks, by adding an implicit contribution of velocity to the momentum equations, is straightforward. However, a simple dimensional analysis reveals that $\mu_\Sigma = \sigma \, \Delta t \, |\nabla \psi|$ represents a viscosity acting in the interface region \citep{Denner2017} and, in conjunction with a Laplacian of velocity $\Delta_\text{s} \vecu$, the second term on the right-hand side of Eq.~\eqref{eq:surfacetensionRaessi} acts as an additional viscous stress term at the interface. As discussed by \citet{Popinet2018}, the resulting increase in dissipation in the vicinity of the interface is responsible for the success of this semi-implicit surface tension formulation. To this end, we demonstrated that an explicit implementation of the additional surface dissipation term in Eq.~\eqref{eq:surfacetensionRaessi} also allows to breach the capillary time-step constraint by up to one order of magnitude \citep{Denner2017}, at the cost of artificially increasing the effective viscous stresses acting in the vicinity of the fluid interface. The implicit treatment of surface tension as part of the proposed algorithm, Eq.~\eqref{eq:csf_imp}, does not introduce an additional viscous stress term at the interface.

We previously presented a coupled implicit algorithm \citep{Denner2015}, similar to the algorithm proposed in Section \ref{sec:numerics}, which also solves the VOF advection equation together with the continuity and momentum equations in a single system of linear equations, and features a semi-implicit treatment of the source term representing surface tension. 
Contrary to the proposed algorithm, in the previous algorithm \citep{Denner2015} the advection terms of the momentum equations \eqref{eq:momentum_disc} and the VOF advection equation \eqref{eq:vof_disc} were linearised with a Picard linearisation, 
\begin{equation}
    \iiint_{V} \frac{\partial u_i \phi}{\partial x_i} \, \text{d}V \approx \sum_f \tilde{\phi}_{f}\imp F_f\rhs, 
    \label{eq:picard}
 \end{equation}
 where $\phi \in \{\vecu, \psi\}$, and the source term representing surface tension only considered the gradient of the colour function in an implicit manner, 
 \begin{equation}
     \mathbf{S}_P\imp \approx \frac{\sigma \kappa_P\rhs}{V_P} \sum_f \overline{\psi}_f\imp \mathbf{n}_f A_f.
     \label{eq:surfacetension2015}
 \end{equation}
 In contrast, in the algorithm proposed in Section \ref{sec:numerics}, a Newton linearisation is applied to linearise the advection terms and the interface curvature is also treated implicitly with respect to the colour function $\psi$.  
 In the previous algorithm \citep{Denner2015}, the advecting velocity was also formulated implicit in the solution variables $\chi \in \{p,u,v,w,\psi\}$, given as
 \begin{equation}
     \begin{split}
         \vartheta_f\imp = \overline{\vecu}_{f}\imp \cdot \mathbf{n}_{f} &-~\hat{d}_f \left[\frac{p_Q\imp-p_P\imp}{\Delta x} - \frac{1}{2} \left(\nabla p_P\rhs + \nabla p_Q\rhs \right) \cdot \mathbf{n}_{f} \right] \\ 
         &+~\hat{d}_f \sigma \left[\overline{\kappa}_f\rhs \frac{\psi_Q\imp-\psi_P\imp}{\Delta x} - \frac{1}{2} \left(\kappa_P\rhs \nabla \psi_P\rhs + \kappa_Q\rhs \nabla \psi_Q\rhs \right)  \cdot \mathbf{n}_{f} \right]\\ 
         &+~\hat{d}_f \frac{\rho}{\Delta t} \left( \vartheta_f\pts - \overline{\vecu}_{f}\pts \cdot \mathbf{n}_{f} \right),
     \end{split}
     \label{eq:mwi_implicit2015}
     \end{equation}
yet, contrary to the formulation proposed in Eq.~\eqref{eq:mwi_implicit}, only the face-based gradients of pressure and colour function were treated implicitly, while the interface curvature and the cell-centred (Gaussian) gradients were deferred. 
As a result of Eqs.~\eqref{eq:surfacetension2015} and \eqref{eq:mwi_implicit2015}, each governing equation has an implicit contribution of the colour function and the continuity equation has implicit contributions with respect to all solution variables $\chi \in \{p,u,v,w,\psi\}$. Nevertheless, the algorithm previously proposed in \citep{Denner2015} does not allow to breach the capillary time-step constraint.

\section{Results}
\label{sec:results}

The ability of the numerical framework proposed in Section \ref{sec:numerics} to breach the capillary time-step constraint is tested and validated using two well-defined test-cases: the Laplace equilibrium of an interface with constant curvature ({\em i.e.}~a circular and a spherical interface) and a standing capillary wave. Both test-cases have frequently been used to scrutinise numerical methods, are governed by surface tension, and analytical solutions are available for comparison. Furthermore, as both considered test-cases yield relatively small interface deformations and the interface curvature is spatially well resolved at all times, these test-cases are not subject to issues associated with the fidelity of the interface transport or the well-posedness of the heights used for the evaluation of the interface curvature, issues that are outside the scope of this study.

\subsection{Laplace equilibrium}
\label{sec:laplace}

A circular or spherical interface subject to surface tension in a quiescent flow is in mechanical equilibrium, with zero velocity in the entire domain and a pressure difference between the outside and the inside of the interface given by the Young-Laplace equation, 
\begin{equation}
    \Delta p = \sigma \kappa.
\end{equation}
However, due to errors associated with the numerical framework and the discrete representation of the interface topology, unphysical parasitic currents arise. For a force-balanced numerical framework, such as the one applied in this study \citep{Denner2014a}, these parasitic currents should decay exponentially due to viscous dissipation and converge to a value commensurate with machine precision or the chosen solver tolerance, with the interface attaining a discrete equilibrium shape \citep{Popinet2009}.

\begin{figure}[t]
    \begin{center}
    \includegraphics[width=\linewidth]{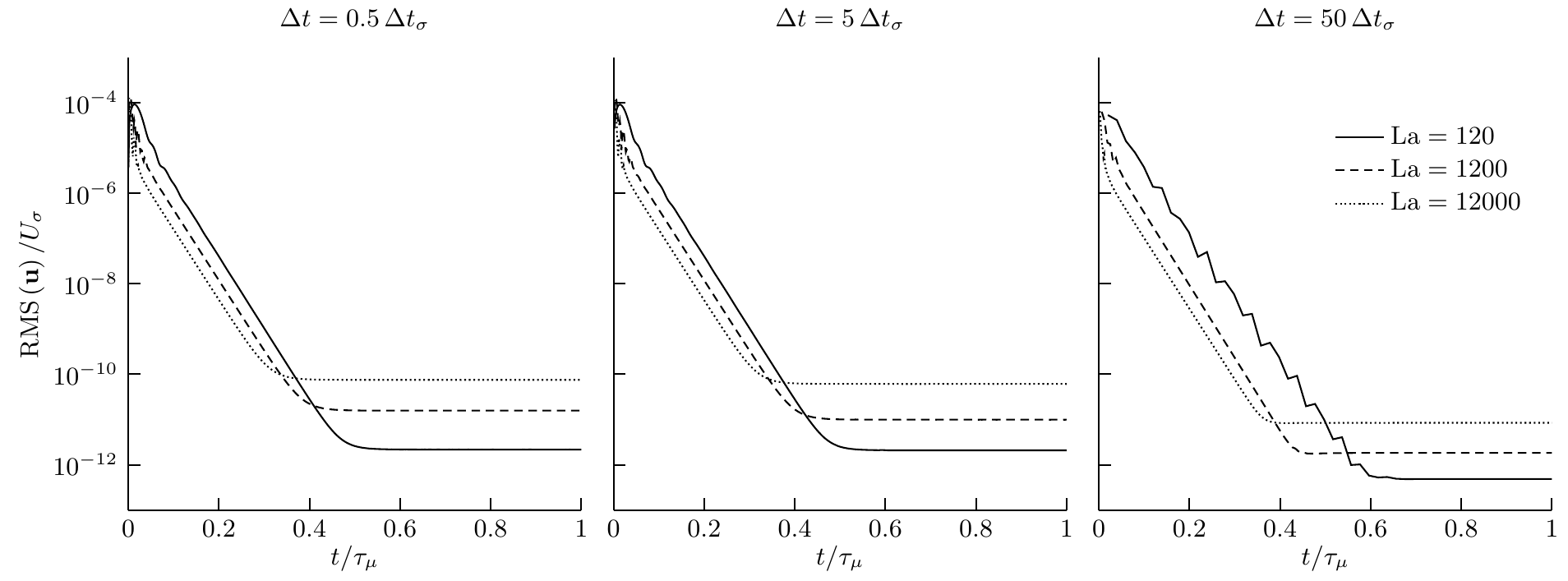}
    \caption{Evolution of the root-mean-square (RMS) of the flow velocity $\mathbf{u}$ of the two-dimensional Laplace equilibrium with Laplace number $\text{La} \in \{120, 1200, 12000\}$, for different time-steps $\Delta t$, normalised by the capillary velocity $U_\sigma = \sqrt{\sigma / (\rho D)}$ and the viscous timescale $\tau_\mu = \rho D^2/\mu$. $\Delta t_\sigma$ refers to the capillary time-step constraint given in Eq.~\eqref{eq:tsigma_Denner}.}
    \label{fig:laplaceeq_2d}
    \end{center}
\end{figure}

\begin{figure}[t]
    \begin{center}
    \includegraphics[width=0.381\textwidth]{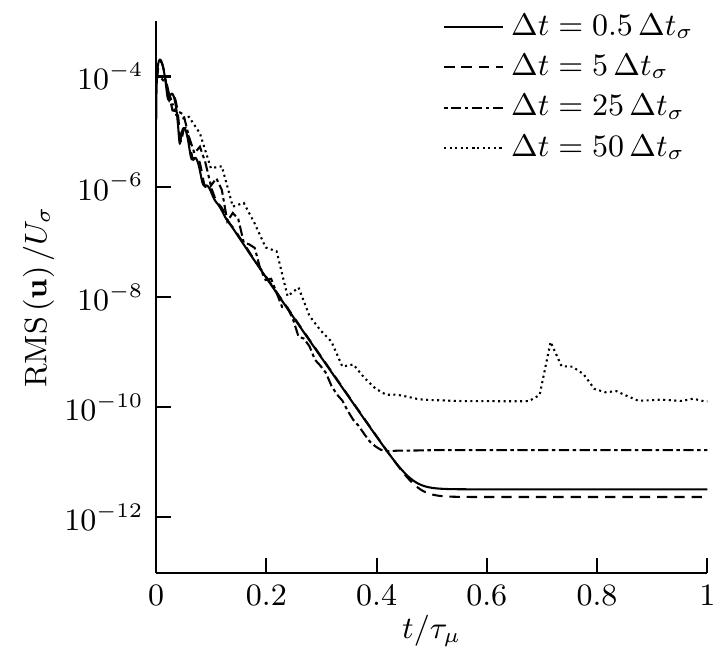}
    \caption{Evolution of the root-mean-square (RMS) of the flow velocity $\mathbf{u}$ of the three-dimensional Laplace equilibrium with Laplace number $\text{La} =120$, for different time-steps $\Delta t$, normalised by the capillary velocity $U_\sigma = \sqrt{\sigma / (\rho D)}$ and the viscous timescale $\tau_\mu = \rho D^2/\mu$. $\Delta t_\sigma$ refers to the capillary time-step constraint given in Eq.~\eqref{eq:tsigma_Denner}.}
    \label{fig:laplaceeq_3d}
    \end{center}
\end{figure}

Following the work of \citet{Popinet2009}, a circular interface with a diameter of $D=0.8 \, \text{m}$ is simulated, with $\rho = 1 \, \text{kg/m}^3$ and $\sigma = 1 \, \text{N/m}$. The viscosity $\mu$ is defined by the considered Laplace number,
\begin{equation}
    \text{La} = \frac{\rho \sigma D}{\mu^2}.
\end{equation}
Exploiting the symmetry of the problem, only one quarter of the circular interface is simulated, situated at the origin of a square domain with edge length $1 \, \text{m}$. The domain is represented by an equidistant Cartesian mesh with $32\times 32$ cells.
Figure \ref{fig:laplaceeq_2d} shows the evolution of the root-mean-square (RMS) of the velocity in the domain for $\text{La} \in \{120, 1200, 12000\}$, simulated with different time-steps $\Delta t$. A stable solution is obtained even if the time-step exceeds the capillary time-step constraint $\Delta t_\sigma$, Eq.~\eqref{eq:tsigma_Denner}, by factor $50$, and the equilibrium velocity, which is of negligible magnitude, is hardly affected.
Considering the same case in three-dimensions, {\em i.e.}~a spherical interface, with $\text{La} = 120$, a similar evolution of the velocity RMS is observed in Figure \ref{fig:laplaceeq_3d}.


Hence, the capillary time-step constraint can be breached for both two- and three-dimensional interfaces, without affecting the discrete balance between surface tension and the flow significantly.

\subsection{Capillary wave}
\label{sec:capillarywave}

\begin{figure}
    \begin{center}
        \includegraphics[scale=0.7]{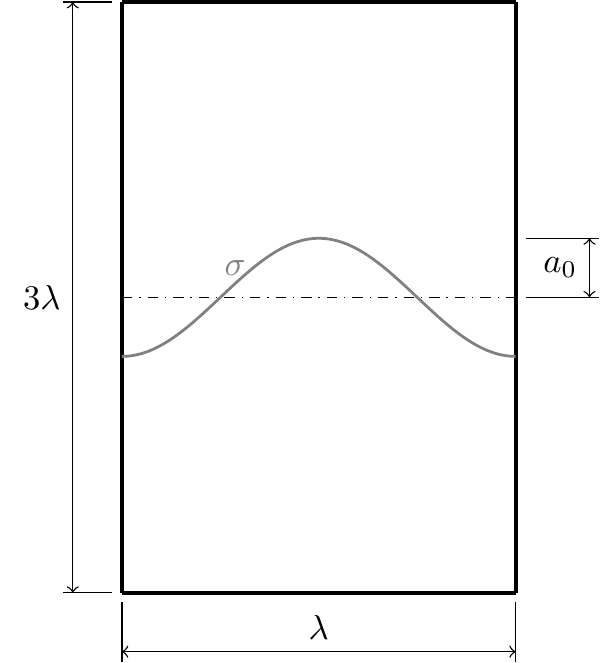}
        \caption{Schematic of the two-dimensional capillary wave with wavelength $\lambda$ and initial amplitude $a_0$.}
        \label{fig:capillarywave_schematic}
    \end{center}
\end{figure}

A single two-dimensional capillary wave is considered to test the fidelity and robustness with which the proposed algorithm can predict surface-tension-driven motion when the capillary time-step constraint is breached.
The motion of the fluid interface and the flow are driven solely by surface tension and, considering a small initial wave amplitude and equal properties of the bulk fluids, an analytical solution to the corresponding initial-value problem is available for comparison \citep{Prosperetti1981}.
The considered capillary wave with wavelength $\lambda=10^{-4} \, \text{m}$ and initial amplitude $a_0 = 0.01 \lambda$ is situated in a domain with dimensions $\lambda \times 3 \lambda$, illustrated in Figure \ref{fig:capillarywave_schematic}, resolved with a mesh spacing of $\Delta x = \lambda/100$. Periodic boundary conditions are applied on the side walls. Both fluids have the same density, $\rho=1 \, \text{kg/m}^3$, and the same viscosity, ranging from $\mu=5 \times 10^{-6} \, \text{Pa s}$ to $\mu=5 \times 10^{-2} \, \text{Pa s}$. The surface tension coefficient is $\sigma=0.01 \, \text{N/m}$. The capillary wave is fully characterised by its critical wavenumber $k_\text{c} \simeq 2 ^{2/3} \sigma (\rho_\text{a}+\rho_\text{b})/(\mu_\text{a}+\mu_\text{b})^2$, above which the oscillation of the capillary wave ceases \citep{Denner2017a}, and its undamped frequency $\omega_0 = \sqrt{\sigma k^3/(\rho_\text{a}+\rho_\text{b})}$, where $k=2 \pi/\lambda$ is the wavenumber.

\begin{figure}[t]
    \includegraphics[width=\linewidth]{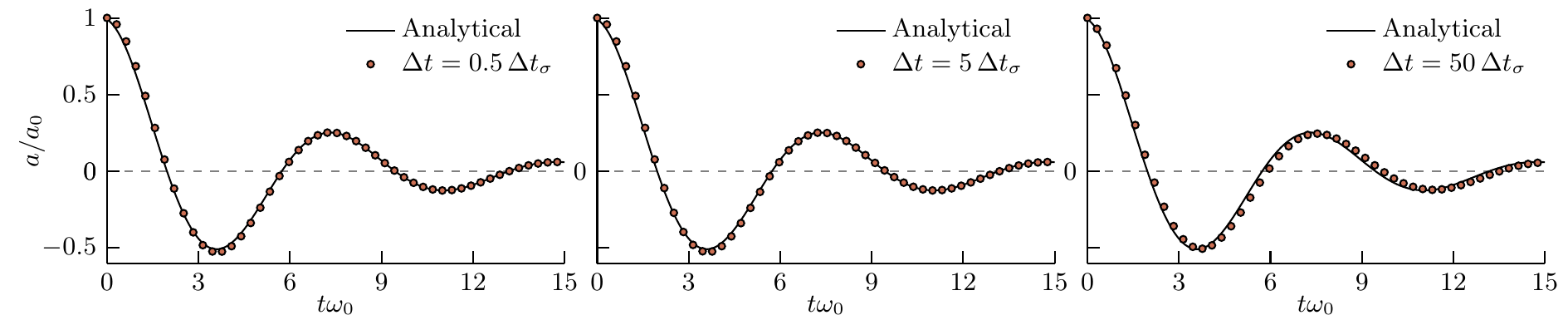}
    \caption{Evolution of the amplitude of a capillary wave with wavelength $\lambda = 202 \, \lambda_\text{c}$ and initial amplitude $a_0=0.01 \lambda$ obtained with different time-steps $\Delta t \in \{0.5,5,50\} \Delta t_\sigma$, where $\Delta t_\sigma$ is the capillary time-step constraint, Eq.~\eqref{eq:tsigma_Denner}. The evolution is presented relative to the undamped frequency $\omega_0$ of the capillary wave and the analytical solution of \citet{Prosperetti1981} is shown as a reference. For the numerical results, each \revB{dot} shows every $100^\text{th}$ time-step for $\Delta t = 0.5 \, \Delta t_\sigma$, every $10^\text{th}$ time-step for $\Delta t = 5 \, \Delta t_\sigma$ and every time-step for $\Delta t = 50 \, \Delta t_\sigma$.}
    \label{fig:capillarywave_400lvc}
\end{figure}

\begin{figure}[t]
    \includegraphics[width=\linewidth]{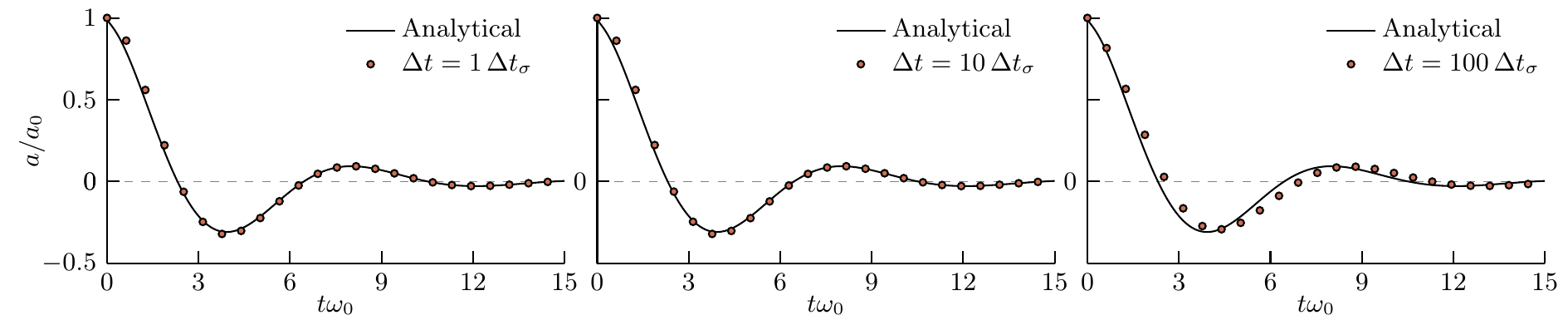}
    \caption{Evolution of the amplitude of a capillary wave with wavelength $\lambda = 12.6 \, \lambda_\text{c}$ and initial amplitude $a_0=0.01 \lambda$ obtained with different time-steps $\Delta t \in \{1,10,100\} \Delta t_\sigma$, where $\Delta t_\sigma$ is the capillary time-step constraint, Eq.~\eqref{eq:tsigma_Denner}. The evolution is presented relative to the undamped frequency $\omega_0$ of the capillary wave and the analytical solution of \citet{Prosperetti1981} is shown as a reference. For the numerical results, each \revB{dot} shows every $100^\text{th}$ time-step for $\Delta t = 1 \, \Delta t_\sigma$, every $10^\text{th}$ time-step for $\Delta t = 10 \, \Delta t_\sigma$ and every time-step for $\Delta t = 100 \, \Delta t_\sigma$.}
    \label{fig:capillarywave_100lvc}
\end{figure}

\begin{figure}[t]
    \begin{center}
    \includegraphics[width=0.38\linewidth]{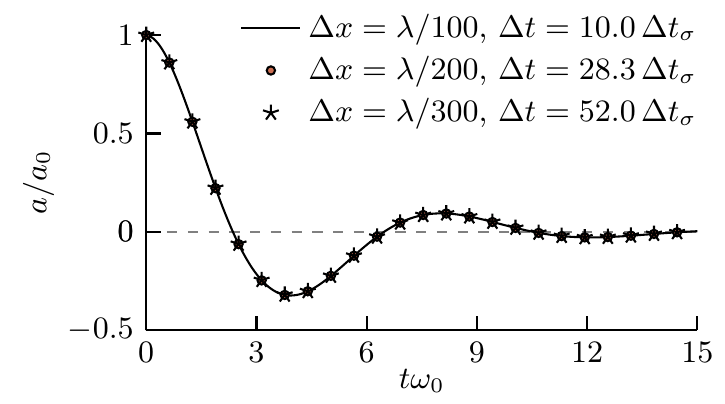}
    \caption{Evolution of the amplitude of a capillary wave with wavelength $\lambda = 12.6 \, \lambda_\text{c}$ and initial amplitude $a_0=0.01 \lambda$. The results are obtained with different mesh resolutions $\Delta x \in \{\lambda/100, \lambda/200, \lambda/300\}$. The same time-step, corresponding to $\Delta t=10 \Delta t_\sigma$ for the mesh with $\Delta x =\lambda/100$, where $\Delta t_\sigma$ is the capillary time-step constraint, Eq.~\eqref{eq:tsigma_Denner}, is applied regardless of the mesh resolution. The evolution is presented relative to the undamped frequency $\omega_0$ of the capillary wave and every $10^\text{th}$ time-step is shown by a mark.}
    \label{fig:capillarywave_dxcomp}
    \end{center}
\end{figure}

Figure \ref{fig:capillarywave_400lvc} shows the evolution of the amplitude of an oscillating capillary wave with a wavelength of $\lambda = 202 \, \lambda_\text{c}$, obtained with different time-steps $\Delta t$. The evolution of the wave amplitude  is predicted accurately compared to the analytical solution, even with $\Delta t = 5 \, \Delta t_\sigma$. Although the numerical algorithm is stable for $\Delta t = 50 \, \Delta t_\sigma$, the result exhibits a visible discrepancy in comparison to the analytical solution. Nevertheless, this discrepancy is to be expected given the rather coarse temporal resolution of the oscillation, whereby each time-step is illustrated by a \revB{dot} in Figure \ref{fig:capillarywave_400lvc}  for $\Delta t = 50 \, \Delta t_\sigma$. Similar observations can be made for the evolution of the amplitude of a relatively shorter capillary wave with a wavelength of $\lambda = 12.6 \, \lambda_\text{c}$ shown in Figure \ref{fig:capillarywave_100lvc}, where a time-step of $\Delta t = 100 \, \Delta t_\sigma$ yields a stable and reasonably accurate result. Changing the mesh resolution with which this capillary wave is resolved, but keeping the applied time-step unchanged, yields virtually identical results, as observed  in Figure \ref{fig:capillarywave_dxcomp}. 

\begin{figure}[t]
    \begin{center}
    \subfloat[All implicit]
    {\includegraphics[width=0.485\linewidth, trim=550pt 650pt 370pt 650pt , clip=true]{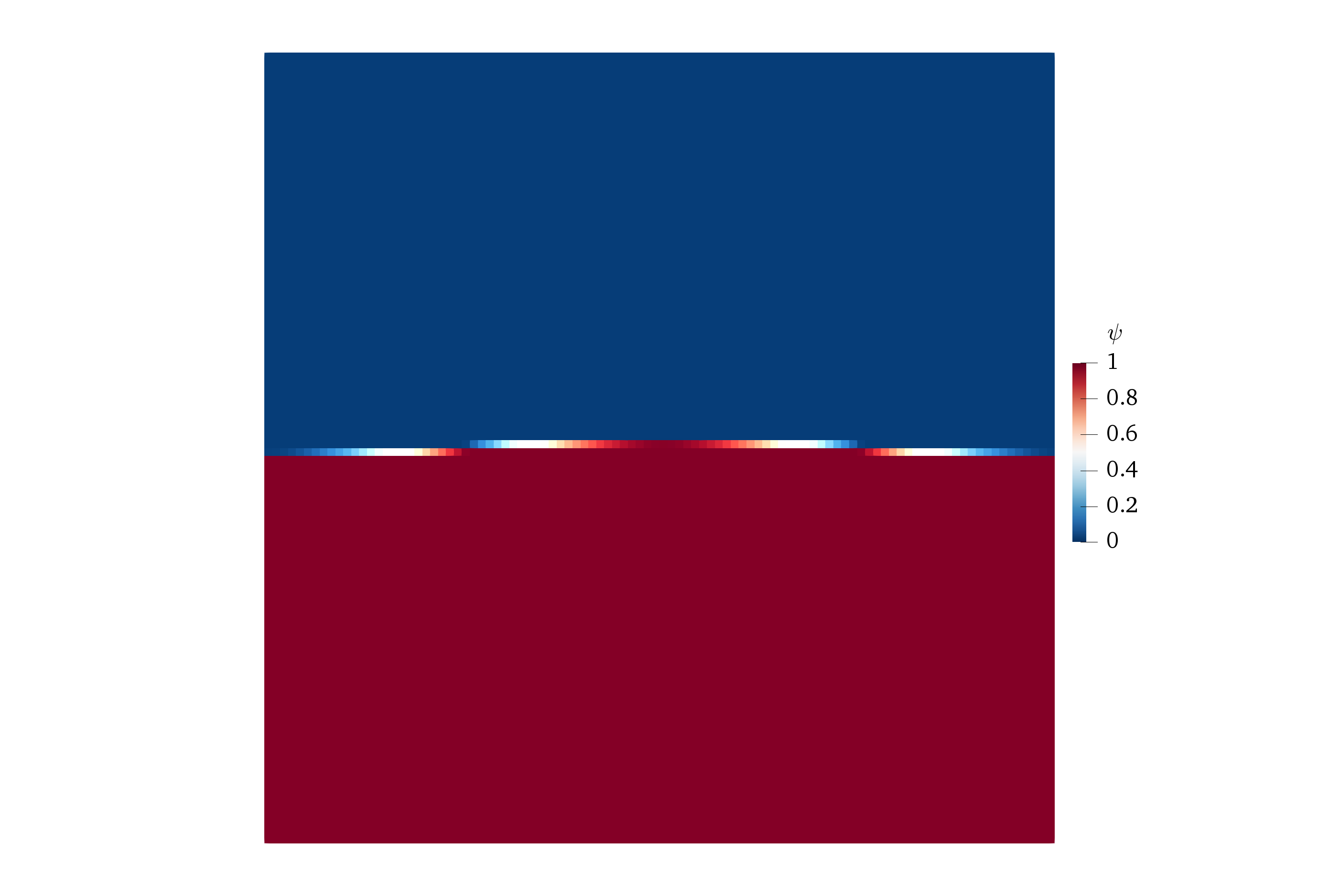}} \quad
    \subfloat[Picard linearisation of the advection terms]
    {\includegraphics[width=0.485\linewidth, trim=550pt 650pt 370pt 650pt , clip=true]{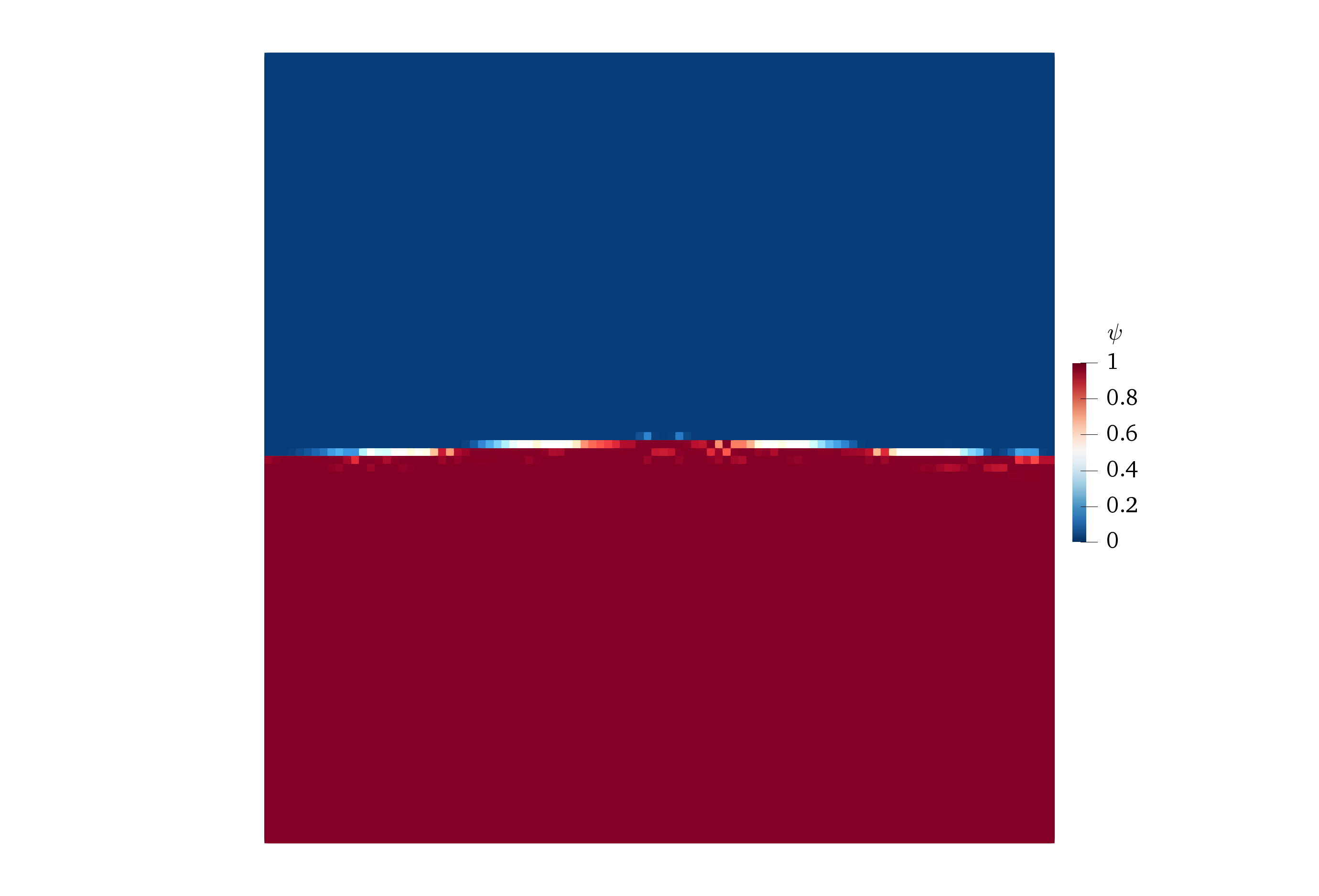}} 
    \\
    \subfloat[Explicit interface curvature]
    {\includegraphics[width=0.485\linewidth, trim=550pt 650pt 370pt 650pt , clip=true]{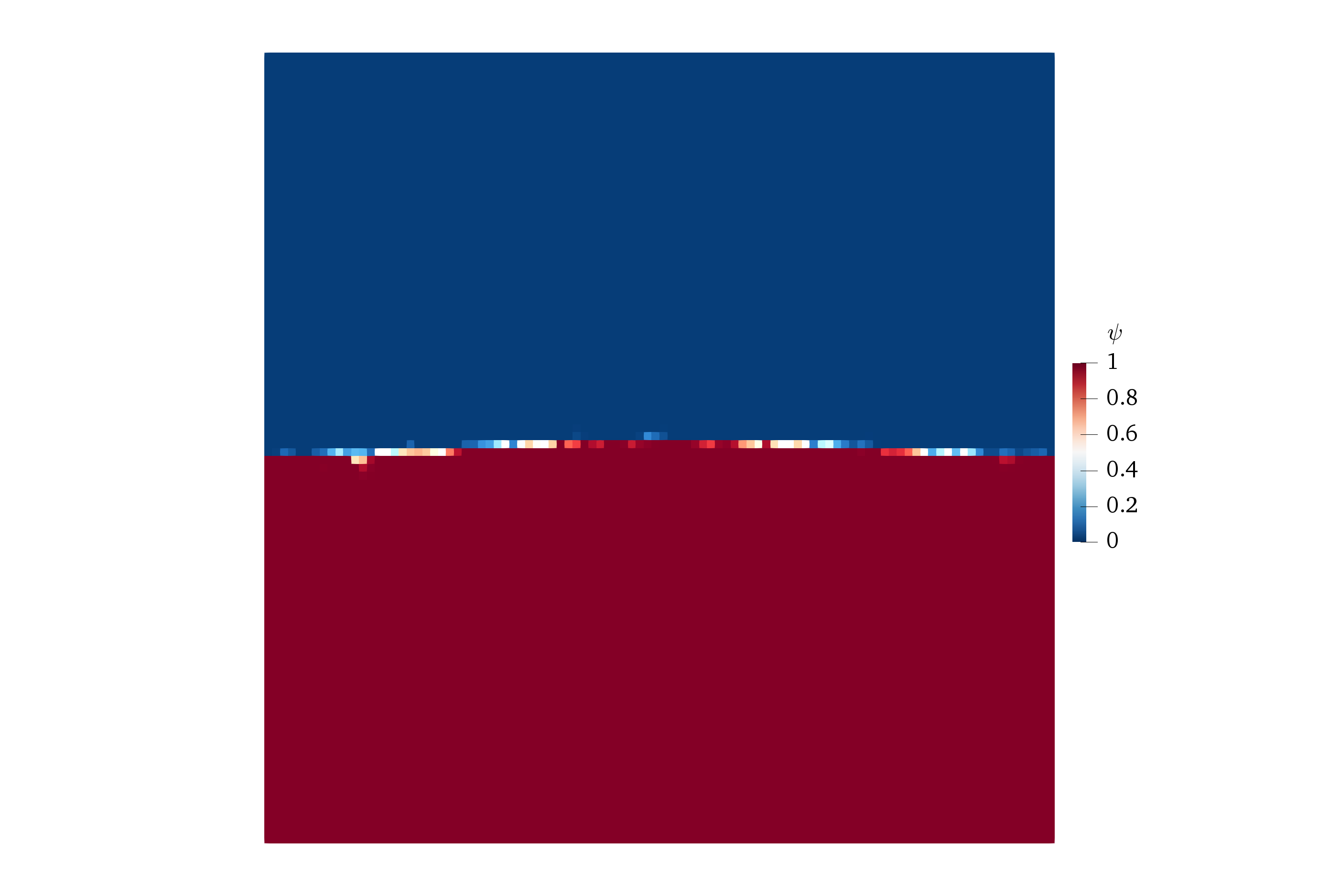}} 
    \quad
    \subfloat[Semi-implicit MWI formulation]
    {\includegraphics[width=0.485\linewidth, trim=550pt 650pt 370pt 650pt , clip=true]{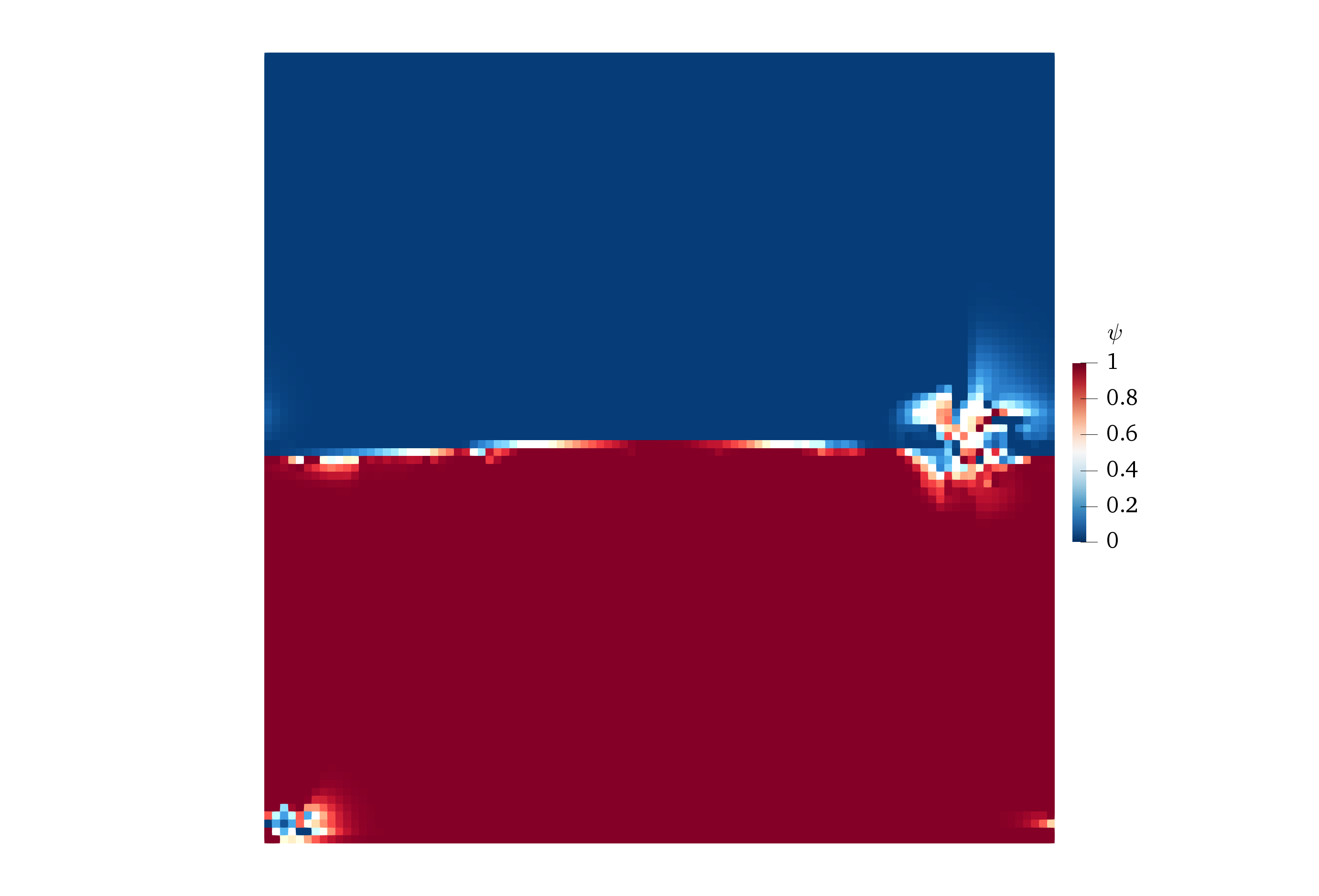}} 
    \caption{Contour plots of the VOF colour function $\psi$ after one time-step of the wave with wavelength $\lambda = 202 \, \lambda_\text{c}$ and initial amplitude $a_0=0.01 \lambda$ obtained with a time-step of $\Delta t = 10 \Delta t_\sigma$ using different modelling assumptions. (a) Shows the result obtained using the proposed {\em all implicit} algorithm; (b) a Picard linearisation, see Eq.~\eqref{eq:picard}, is applied for the linearisation of the advection terms of the momentum and VOF advection equations; (c) the interface curvature is treated explicitly, with the source term representing surface tension given by Eq.~\eqref{eq:surfacetension2015}; (d) the semi-implicit MWI formulation given in Eq.~\eqref{eq:mwi_implicit2015} is applied, but with the proposed implicit treatment of the interface curvature.}
    \label{fig:capillarywave_contourplot}
    \end{center}
\end{figure}

Figure \ref{fig:capillarywave_contourplot} shows contour plots of the colour function $\psi$ after a single time-step with $\Delta t=10 \Delta t_\sigma$, using different modelling assumptions. Neglecting any of the three main implicit extensions proposed in comparison to our previously presented coupled algorithm \citep{Denner2015}, as detailed in Section \ref{sec:comparisonprevious}, yields an unphysical interface topology even after a single time-step, if the capillary time-step is breached. Only the proposed algorithm in which all contributions of the solution variables $\chi \in \{p,u,v,w,\psi\}$ are treated implicitly yields a stable result for time-steps exceeding the capillary time-step constraint.

\section{Revised time-step constraint}
Although the proposed algorithm exhibits a favourable behaviour for time-steps exceeding the capillary time-step constraint, the time-step that yields a stable solution is still limited. We presume that this time-step limitation is associated with the particular discretisation and linearisation chosen to cast the governing equations in a form amenable to numerical analysis using linear algebra.

Because the governing equations are linearised, we consider linear stability analysis to analyse the stability  of the linearised system of governing equations.
Based on a linear stability analysis under the assumption of an interface perturbation with small amplitude and sufficiently small Reynolds number, \citet{Galusinski2008} proposed a maximum time-step $\Delta t_\star$ for a stable solution of surface-tension-driven flows of
\begin{equation}
    \Delta t_\star^2 - c_2 \, \frac{{\mu} \, \Delta x}{\sigma} \Delta t_\star - c_1 \, \frac{{\rho} \, \Delta x^3}{\sigma}=0,
    \label{eq:dtPolynom_Galusinski}
\end{equation}
where $c_1$ and $c_2$ are constants.
With the wavelength of the shortest spatially resolved capillary waves being $\lambda_\sigma = 2 \Delta x$, and assuming\footnote{\citet{Galusinski2008} did neither specify nor discuss how the density and viscosity of the two-phase system are defined in their linear stability analysis.} $\hat{\mu} = \mu_\text{a} + \mu_\text{b}$ and $\hat{\rho} = \rho_\text{a} + \rho_\text{b}$, we reformulate Eq.~\eqref{eq:dtPolynom_Galusinski} as
\begin{equation}
    \Delta t_\star^2 - a_2 \, \frac{\hat{\mu} \, \lambda_\sigma}{\sigma} \Delta t_\star - a_1 \, \frac{\hat{\rho} \, \lambda_\sigma^3}{\sigma}=0,
    \label{eq:dtPolynom_GalusinskiNew}
\end{equation}
or, by inserting the capillary timescale $\tau_\sigma = \sqrt{\hat{\rho}  \lambda_\sigma^3/\sigma}$ and the viscocapillary timescale $\tau_\text{vc} = \hat \mu \lambda_\sigma /\sigma$ \citep{Castrejon-Pita2015},
\begin{equation}
    \Delta t_\star^2 - a_2 \, \tau_\text{vc} \, \Delta t_\star - a_1 \, \tau_\sigma^2 =0.
    \label{eq:dtPolynom_GalusinskiNew_short}
\end{equation}
The maximum time-step $\Delta t_\star$ follows as the positive root of Eq.~\eqref{eq:dtPolynom_GalusinskiNew_short},
\begin{equation}
    \Delta t_\star = \frac{a_2 \, \tau_\text{vc} + \sqrt{a_2^2 \, \tau_\text{vc}^2 + 4 \, a_1 \, \tau_\sigma^2 }}{2}.
    \label{eq:dt_Galusinski}
\end{equation}
This suggests that the maximum applicable time-step is proportional to the capillary timescale $\tau_\sigma$ for small Ohnesorge numbers with respect to $\lambda_\sigma$, 
\begin{equation}
    \text{Oh} = \frac{\tau_\text{vc}}{\tau_\sigma} = \frac{\hat{\mu}}{\sqrt{\hat{\rho} \sigma \lambda_\sigma}},
    \label{eq:Oh}
\end{equation} 
where surface tension dominates. In contrast, a maximum time-step proportional to the viscocapillary timescale $\tau_\text{vc}$ is relevant for large $\text{Oh}$, where both viscosity and surface tension govern the interface motion. Note that the capillary time-step constraint $\Delta t_\sigma$, as presented in Eq.~\eqref{eq:tsigma_Denner}, associated with an explicit treatment of surface tension is recovered for $a_1 = (16\pi)^{-1}$ and $a_2=0$.

\begin{figure}[t]
    \begin{center}
    \subfloat[Laplace equilibrium]
    {\includegraphics[width=0.43\linewidth]{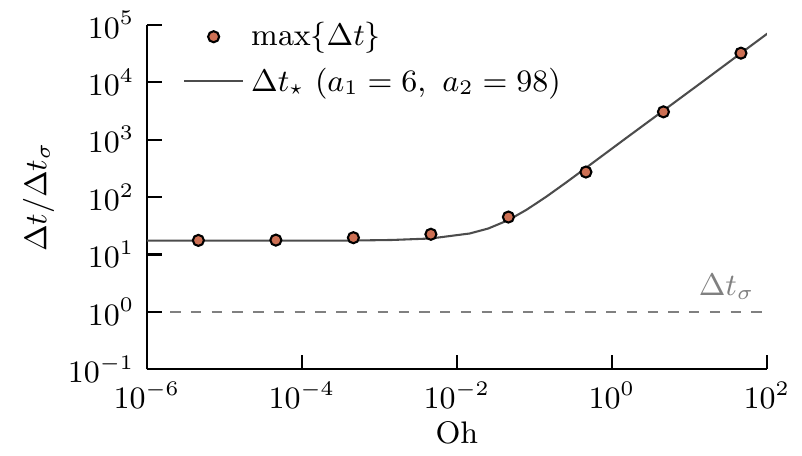} \label{fig:laplace_newdt}}\quad
    \subfloat[Capillary wave]
    {\includegraphics[width=0.43\linewidth]{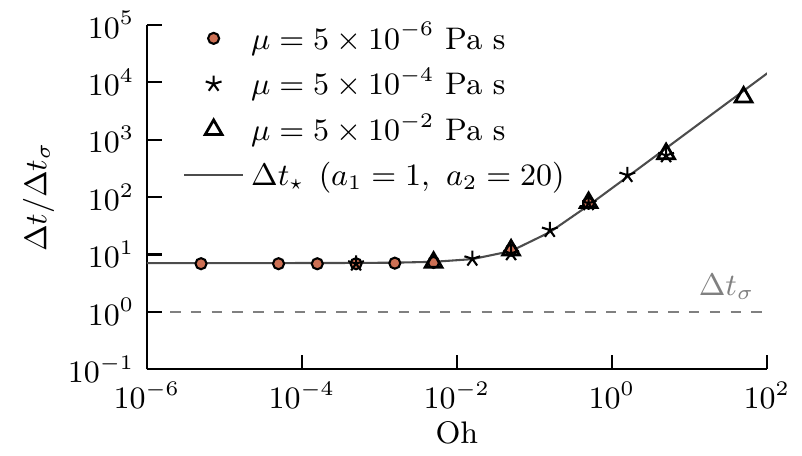} \label{fig:capillarywave_newdt_sigma}}
    \caption{Maximum applicable time-step $\Delta t$, normalised by the capillary time-step constraint $\Delta t_\sigma$ defined in Eq.~\eqref{eq:tsigma_Denner}, as a function of the Ohnesorge number $\text{Oh}$ defined in Eq.~\eqref{eq:Oh}, for (a) the Laplace equilibrium of Section \ref{sec:laplace} and (b) the capillary wave (with three different viscosities) of Section \ref{sec:capillarywave}. The approximate maximum time-step $\Delta t_\star$ using Eq.~\eqref{eq:dt_Galusinski} is shown with $a_1=6$ and $a_2=98$ in (a) and with $a_1=1$ and $a_2=20$ in (b), \revA{where suitable values for $a_1$ and $a_2$ are approximated}.}
    \label{fig:newdt}
    \end{center}
\end{figure}

Figure \ref{fig:newdt} shows the maximum time-step that yields a stable result for the Laplace equilibrium (Section \ref{sec:laplace}) and the capillary wave (Section \ref{sec:capillarywave}), normalised by the capillary time-step constraint $\Delta t_\sigma$ given by Eq.~\eqref{eq:tsigma_Denner}, for different Ohnesorge numbers as defined in Eq.~\eqref{eq:Oh}.
For both cases, the maximum applicable time-step exceeds the capillary time-step constraint $\Delta t_\sigma$ and the revised time-step constraint presented in Eq.~\eqref{eq:dt_Galusinski} is in remarkable agreement with the maximum time-step over the considered eight orders of magnitude of the Ohnesorge number. 

The results presented in Figure \ref{fig:newdt} further indicate that the coefficients $a_1$ and $a_2$ are case dependent; $a_1 \approx 6$ and $a_2 \approx 98$ for the Laplace equilibrium, whereas $a_1 \approx 1$ and $a_2 \approx 20$ for the capillary wave. More generally, the results suggest a maximum applicable time-step of $\mathcal{O} (\tau_\sigma)$ in the surface-tension-dominated regime ($\text{Oh} \ll 0.01$) and of $\mathcal{O} (10 \,  \tau_\text{vc}) - \mathcal{O} (100 \,  \tau_\text{vc})$ in the viscocapillary regime ($\text{Oh} \gg 0.01$). 
\revA{We are currently not aware of a method to estimate precise values for $a_1$ and $a_2$ from first principles or based on the discretisation of the governing equations. However, given that the revised time-step constraint $\Delta t_\star$ is given by a second-order polynomial, the coefficients can be approximated for a given case for all practically relevant Ohnesorge numbers with only two results, one for $\mathrm{Oh} \ll 0.01$ and one for $\mathrm{Oh} \gg 0.01$.}

\section{Conclusions}
\label{sec:conclusions}

The capillary time-step constraint, first formulated by \citet{Brackbill1992}, presents a severe impediment to the performance of most interfacial flow simulations with surface tension. Breaching or even eliminating the capillary time-step constraint is generally thought to be possible with an implicit treatment of surface tension. However, previous work in this direction using interface capturing methods has either been unsuccessful \citep{Denner2015} or has introduced an additional viscous dissipation term \citep{Hysing2006, Raessi2009, Denner2017}. This led us to conclude that it is not possible to breach the capillary time-step constraint with interface capturing methods \citep{Denner2015}, a conclusion that was \revB{(correctly)} met with scepticism \citep{Popinet2018}. 

In this study, we have presented a fully-coupled pressure-based algorithm, based on a second-order finite-volume discretisation, featuring an implicit VOF method and an implicit linearised treatment of surface tension. Three implementation principles are at the heart of this algorithm: (i) making all governing equations implicitly dependent on pressure, velocity and the colour function (through the momentum-weighted interpolation), (ii) linearising all nonlinear terms with a Newton linearisation, and (iii) treating every term involving pressure, velocity or the colour function implicitly. The ensuing system of discretised linear equations, which includes the continuity, momentum and VOF advection equations, is then solved simultaneously for pressure, velocity and the colour function. We have shown that this algorithm is able to breach the capillary time-step constraint; hence, interface capturing methods are indeed able to breach the capillary time-step constraint, which proves our early conclusion in \citep{Denner2015} to be incorrect. The presented results further indicate that the proposed algorithm features the minimum level of implicitness required for breaching the capillary time-step constraint.  

However, this study also highlights the limitation of the proposed approach; by how much the capillary time-step constraint can be breached depends on the fluid properties as well as the considered case. To this end, the maximum time-step that yields a stable solution is described accurately by a revised time-step constraint based on a linear stability analysis previously proposed by \citet{Galusinski2008}. As a result, the maximum time-step depends on the surface tension coefficient, density and viscosity, as well as two coefficients that are case dependent. Nevertheless, stable results with time-steps larger than the capillary time-step constraint have been obtained for all simulations considered. 

With this study, we provide a proof-of-concept for a fully-coupled algorithm and an implicit treatment of surface tension, based on an interface capturing method, that allows to breach the capillary time-step constraint. While the primary aim of breaching the capillary time-step constraint has been achieved, additional work is required to make such an algorithm applicable to solve problems relevant in practice and further exploit its benefits. For instance, 
\revA{in this proof-of-concept we have only considered density and viscosity ratios equal to unity. Making the proposed algorithm applicable to flows with density and viscosity ratios as they occur in practice, while simultaneously allowing to breach the capillary time-step, may additionally require to treat the density and viscosity in a semi-implicit fashion as a function of the colour function.
Furthermore,} in the employed algebraic VOF method, the advection of the colour function is based on the CICSAM scheme, which is known for requiring very small time-steps to retain a sharp interface \citep{Gopala2008} and, thus, stands in opposition to  maximising the applied time-step. The colour function in the vicinity of the interface is, hence, smeared quickly if the interface moves significantly and, as a consequence, the height-function method becomes ill-posed. Contemporary interface advection schemes, such as THINC \citep{Xiao2011}, would perhaps be better suited to maximise the time-step effectively and robustly. Also, the proposed numerical framework is not limited to implicit algebraic VOF methods, but may be based upon an implicit phase-field or level-set method, and it remains to be determined which interface capturing method is best suited to exploit the benefits of the proposed fully-coupled approach. \revB{Even a geometric VOF method \citep[\textit{e.g.}][]{Youngs1982} could, in principle, be used in conjunction with the proposed numerical framework, on the condition that it can be implemented implicitly and solved in a coupled system simultaneous with the governing equations.}

\section*{Acknowledgements}
This research was funded by the Deutsche Forschungsgemeinschaft (DFG, German Research Foundation), grant numbers 452916560 and 458610925.

\end{document}